\documentclass[a4paper,twocolumn,11pt,unpublished]{quantumarticle}
\pdfoutput=1

\usepackage[utf8]{inputenc}
\usepackage[T1]{fontenc}
\usepackage[english]{babel}
\usepackage[numbers,sort&compress]{natbib}

\usepackage{tikz}
\usetikzlibrary{arrows.meta}

\usepackage{amsmath,amssymb,mathtools}
\usepackage[nointegrals]{wasysym}
\usepackage{graphicx}
\usepackage{romannum}
\usepackage{booktabs}
\usepackage[strings]{underscore}
\usepackage{array}
\usepackage{xcolor}
\usepackage{microtype}
\usepackage{ragged2e}
\usepackage{adjustbox}
\usepackage{hyperref}
\usepackage[capitalize,nameinlink]{cleveref}
\crefname{figure}{Fig.}{Figs.}
\Crefname{figure}{Fig.}{Figs.}

\newcommand{\Psucc}{\ensuremath{P_{\mathrm{succ}}}}
\newcommand{\Vcost}{\ensuremath{V}}
\newcommand{\Vexp}{\ensuremath{V_{\mathrm{exp}}}}

\newcommand{\ket}[1]{\ensuremath{\left|#1\right\rangle}}
\newcommand{\Tstate}{\ensuremath{\ket{T}}}
\newcommand{\epsout}{\ensuremath{\epsilon_{\mathrm{out}}}}

\newcommand{\Qpeak}{\ensuremath{Q_{\mathrm{peak}}}}
\newcolumntype{L}[1]{>{\raggedright\arraybackslash}p{#1}}

\begin{document}

\title{A Resource Comparison of Logical T-State Preparation}

\author{Jianshuo Gao}
\affiliation{School of Physics, Peking University, Beijing 100871, China}

\author{Xiao Yuan}
\email{xiaoyuan@pku.edu.cn}
\affiliation{Center on Frontiers of Computing Studies, School of Computer Science, Peking University, Beijing 100871, China}

\author{Yuan Yao}
\email{yuan.yao@pku.edu.cn}
\affiliation{Center on Frontiers of Computing Studies, School of Computer Science, Peking University, Beijing 100871, China}

\date{April 27, 2026}

\begin{abstract}
Logical $T$-state preparation is a major overhead source in fault-tolerant architectures built from stabilizer operations. Existing protocols, however, are reported under different code families, noise models, postselection rules, and cost conventions, making direct comparison difficult. We compare three representative preparation routes, namely magic-state distillation, magic-state cultivation, and code-switching, using currently available results. Rather than reducing heterogeneous data to a single cost metric, we retain source-native cost units and record output error, single-attempt cost, expected cost per accepted output, footprint, latency, and reporting completeness for each configuration. Within the current dataset, distillation reaches the lowest output-error regime; code-switching achieves the lowest reported single-attempt cost and the smallest explicit footprint among the compatible rows; and recent $\mathbb{RP}^2$ cultivation results add low-cost cultivation points in the $10^{-6}$--$10^{-9}$ output-error range. As a simple algorithm-level case study, we also examine the reported preparation routes under an error budget motivated by Shor’s factoring algorithm, in order to relate single-state preparation costs to full-workload requirements. The resulting comparison clarifies the trade-offs currently supported across the literature, while remaining bounded by the conventions and coverage of the underlying papers.
\end{abstract}

\maketitle

\section{Introduction}\label{sec:introduction}

Large-scale quantum algorithms motivate the development of fault-tolerant architectures capable of executing long computations reliably on encoded qubits~\cite{feynman1982simulating,shor1994algorithms,preskill2018nisq,shor1996fault,aharonov2008fault,gottesman2010qec}. Within many such architectures, particularly those based on stabilizer codes, Clifford operations are comparatively natural, whereas a non-Clifford resource remains necessary for universality. At the circuit level, a common way to supply this resource is to prepare a $\Tstate$ magic-state and consume it through a teleportation-based state-injection procedure to realize a $T$ gate~\cite{gottesman1999teleportation,bravyi2005universal,campbell2017roads}. 

In a fault-tolerant implementation, the injection procedure is carried out at the logical level. Quantum information is protected by logical qubits, so the required resource is a high-fidelity logical $\Tstate$. A direct route, for example in a surface-code setting~\cite{fowler2012surface,litinski2019magic}, injects a physical or small-code $\Tstate$ into an encoded patch and then uses stabilizer measurements and repeated syndrome extraction to grow, verify, or maintain the patch. Such directly prepared logical $\Tstate$s are generally noisy, so additional preparation procedures are needed to suppress the output error while controlling the associated space-time overhead.

Because high-fidelity $\Tstate$ preparation typically requires substantial qubit, time, and routing overhead, while large-scale algorithms consume such states in very large numbers, the rate at which logical $\Tstate$s can be supplied often becomes a dominant bottleneck in fault-tolerant architectures, strongly affecting overall space-time cost, layout, and computational throughput.~\cite{ogorman2017realistic,litinski2019magic,litinski2019notcostly,gidney2021factor}.

Several preparation strategies have therefore been proposed. Magic-state distillation converts many noisy resource states into fewer higher-fidelity outputs and remains the most established route to very low logical error rates~\cite{bravyi2005universal,bravyi2012low,haah2017lowspace,jones2013multilevel,hastings2018sublog}. Code-switching and gauge-fixing approaches instead exploit code families in which non-Clifford gates or resource states are easier to access, at the cost of converting quantum information between encodings~\cite{anderson2014fault,bombin2015gauge,beverland2021cost,daguerre2025code}. More recently, magic-state cultivation protocols have sought to grow and verify a single encoded magic-state through staged code evolution, reducing reliance on large multi-copy distillation factories~\cite{gidney2024cultivation,chen2026efficient,vaknin2025efficient,claes2025cultivating}.

Several prior works have already compared resource costs for particular preparation routes. For example, Gidney, Shutty, and Jones place magic-state cultivation against earlier space-time-volume estimates and show that cultivation can substantially reduce the qubit-round cost in an intermediate-output-error regime~\cite{gidney2024cultivation}. However, such comparisons are still tied to the assumptions and cost conventions of the original papers: they do not provide a single common framework covering distillation, code-switching, and cultivation simultaneously. The underlying studies differ in code family, noise model, decoder assumption, postselection rule, time baseline, and resource unit. The difficulty is therefore not that such protocols cannot be compared, but that the reported costs are not always defined on the same basis. Moreover, the reported cost units are not identical across papers. Some estimates are based on physical-qubit footprints and code cycles, others are given in logical-level space-time units, and postselected protocols may report either a single-attempt cost or an expected cost per accepted output. A cross-study ranking can therefore be misleading if these quantities are compressed into a single scalar without preserving the assumptions under which they were reported. Prior work has addressed this issue more directly by enforcing common assumptions for a narrower comparison. In particular, Beverland, Kubica, and Svore directly compared magic-state distillation and code-switching under a unified circuit-noise setting~\cite{beverland2021cost}. We use that work as a methodological reference for the need to separate shared-assumption comparisons from heterogeneous literature summaries; it is not included as a numerical row in the present dataset.

Our aim is instead to compare the existing evidence on its own terms. We compile quantitative results for logical $\Tstate$ preparation across distillation, code-switching, and cultivation, recording for each configuration the output error $\epsout$, single-attempt cost $\Vcost$, expected cost $\Vexp$, peak footprint $\Qpeak$, and latency $D$ when these quantities are explicitly reported or can be reconstructed from the stated data. Cost metrics are retained in their original units unless a clear conversion is available. Although this limits the strength of direct cross-protocol numerical claims, it preserves the reporting basis of each result and allows robust conclusions to be identified across the current literature.

The main contributions of this work are as follows. First, we place distillation, code-switching, and cultivation within a single comparative framework, enabling a direct view of three protocol families that are usually assessed separately. Second, we develop a comparison methodology tailored to heterogeneous literature, one that retains source-native cost conventions, exposes missing fields explicitly, and distinguishes comparisons made under shared assumptions from broader cross-paper summaries. Third, this framework makes it possible to identify the trade-off landscape currently supported by existing evidence, including the deepest low-error regime reached by distillation, the low-cost and small-footprint region occupied by code-switching, and the recent improvement of cultivation in the $10^{-6}$–$10^{-9}$ infidelity regime. Finally, we use a simple Shor-style error-budget criterion to illustrate the difference between single-state efficiency and algorithm-level sufficiency.

The remainder of the paper is organized as follows. \Cref{sec:preliminaries} reviews the three preparation mechanisms. \Cref{sec:method} defines the fields and comparison rules. \Cref{sec:results} presents the regime map and the family comparison. \Cref{sec:shor_case} gives an illustrative algorithm-level error-budget check. \Cref{sec:limitations,sec:conclusion} state the limitations and conclusions.

\section{Preliminaries}\label{sec:preliminaries}
In fault-tolerant quantum computing, universal logical computation is typically achieved by supplementing Clifford operations with a non-Clifford gate, most commonly the logical $T$ gate. In many architectures, however, the logical $T$ gate is not implemented directly. Instead, it is realized by consuming a logical magic-state,
\begin{align}
    \ket{T}=\frac{1}{\sqrt{2}}\left(\ket{0}+e^{i\pi/4}\ket{1}\right),
\end{align}
through state-injection, usually via a gate-teleportation circuit~\cite{gottesman1999teleportation,zhou2000methodology,bravyi2005universal}. This construction makes logical $\Tstate$ preparation a central ingredient of universal fault-tolerant computation.

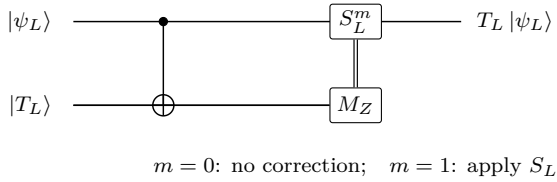
\begin{figure}[t]
\centering
\begin{tikzpicture}[
    x=0.82cm,y=0.82cm,
    every node/.style={font=\scriptsize},
    wire/.style={line width=0.55pt},
    ccwire/.style={line width=0.45pt,double,double distance=0.8pt},
    gate/.style={draw,rounded corners=1pt,minimum width=0.68cm,minimum height=0.46cm,inner sep=1.2pt,fill=white}
]

\node[anchor=east] at (-0.18,1.35) {$\ket{\psi_L}$};
\draw[wire] (0,1.35) -- (6.25,1.35);
\node[anchor=west] at (6.34,1.35) {$T_L\ket{\psi_L}$};

\node[gate] (mz) at (4.55,0) {$M_Z$};
\draw[wire] (0,0) -- (mz.west);

\node[anchor=east] at (-0.18,0) {$\ket{T_L}$};
\draw[wire] (0,0) -- (mz.west);

\filldraw (1.45,1.35) circle (1.55pt);
\draw[wire] (1.45,1.35) -- (1.45,0);
\draw[wire] (1.45,0) circle (0.17);
\draw[wire] (1.28,0) -- (1.62,0);
\draw[wire] (1.45,-0.17) -- (1.45,0.17);

\node[gate] (s) at (4.55,1.35) {$S_L^m$};
\draw[ccwire] (4.55,0.29) -- (s.south);

\node[anchor=north,align=center,font=\scriptsize] at (4.55,-0.68)
    {$m=0$: no correction;\quad $m=1$: apply $S_L$};
\end{tikzpicture}
\caption{Logical-level magic-state-injection gadget for implementing a logical $T$ gate. The data block controls a logical CNOT into a prepared logical magic-state, the magic-state block is measured in the logical $Z$ basis, and the measurement result $m\in\{0,1\}$ controls the Clifford correction $S_L^m$. With this convention, the output is $T_L\ket{\psi_L}$ up to an irrelevant global phase~\cite{gottesman1999teleportation,bravyi2005universal}.}
\label{fig:logical_injection}
\end{figure}

The difficulty arises because logical Clifford operations are often supported relatively naturally by fault-tolerant primitives such as transversal gates, lattice surgery, or code deformation, whereas the logical $T$ gate is generally not available in such a form. More broadly, the structure of protected gate sets is constrained by general no-go and locality results~\cite{eastin2009restrictions,bravyi2013classification,campbell2017roads}. As a result, preparing high-fidelity logical $\Tstate$s typically incurs substantial overhead and becomes a major resource bottleneck in scalable fault-tolerant architectures.

\begin{figure*}[t]
\centering
\resizebox{!}{0.76\textheight}{%
\input{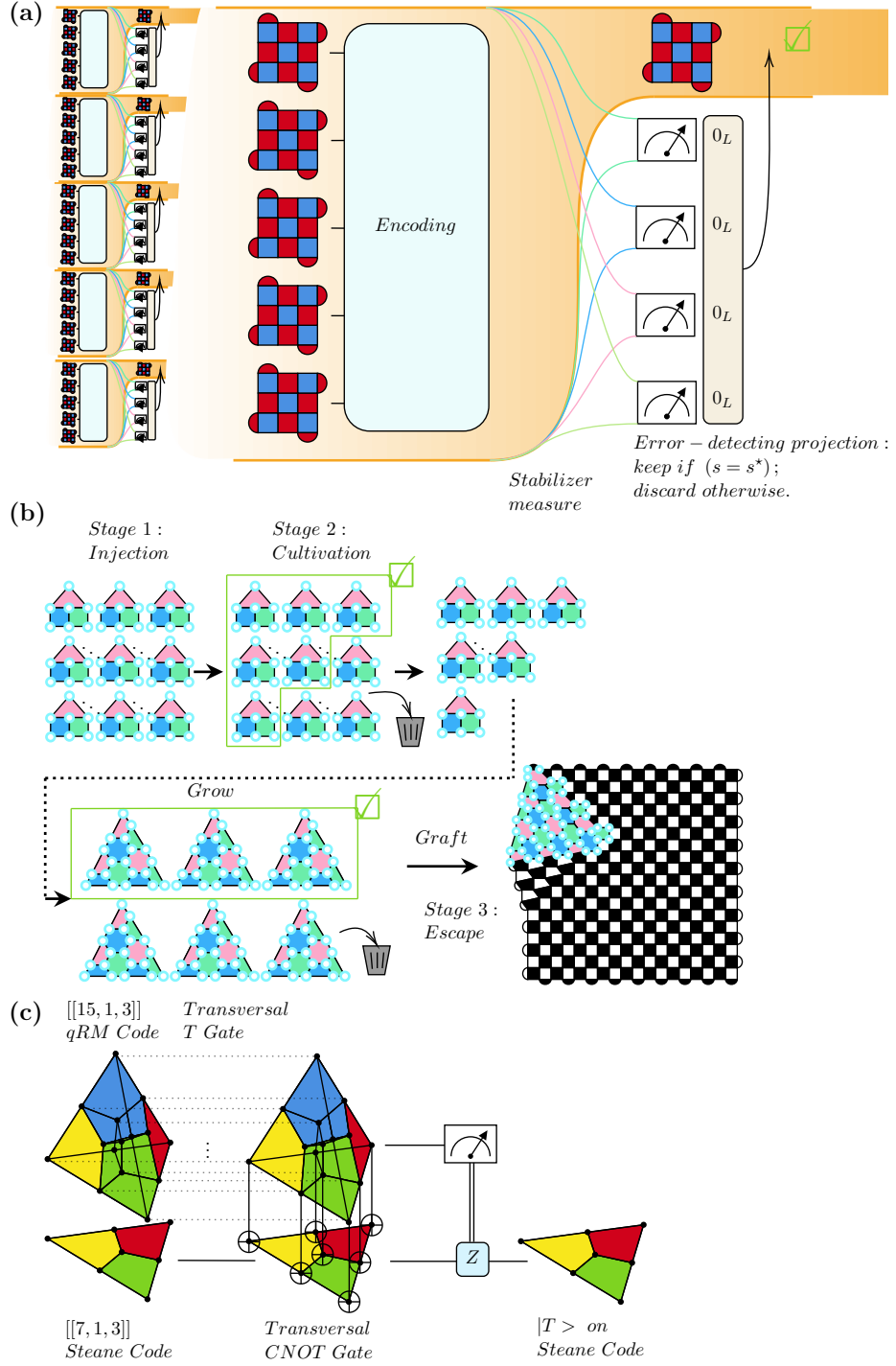}%
}
\caption{Representative logical $\Tstate$-preparation routes. 
(a) Distillation processes several noisy logical magic-states using a Clifford check circuit, measures stabilizer checks to obtain a syndrome vector $s$, and keeps the run only when $s=s^\star$; otherwise the run is discarded ~\cite{bravyi2005universal,bravyi2012low,litinski2019magic}.
(b) Cultivation injects a seed into a distance-$3$ color code, grows and verifies it, and grafts accepted states into a larger surface-code patch ~\cite{gidney2024cultivation,chen2026efficient}.
(c) Code-switching prepares a logical $\Tstate$ in the
$\left[\!\left[15,1,3\right]\!\right]$ qRM code via transversal $T$, transfers it to the $\left[\!\left[7,1,3\right]\!\right]$ Steane code using transversal CNOT, and postselects on the qRM measurement outcomes ~\cite{anderson2014fault,bombin2015gauge,daguerre2025code}.}
\label{fig:abc_schematic}
\end{figure*}

This section reviews three representative approaches to logical magic-state preparation: magic-state distillation, magic-state cultivation, and code-switching. 
The three mechanisms are summarized schematically in Fig.~\ref{fig:abc_schematic}, with panels (a)--(c) corresponding respectively to distillation, cultivation, and code-switching.

Figure~\ref{fig:abc_schematic} summarizes the protocols used below: distillation is a many-to-one error-detection routine, cultivation grows and verifies one developing encoded state, and code-switching prepares or transfers the resource through a code family with more favorable protected non-Clifford structure. The following subsections give the details used in the comparison.

\subsection{Magic-state distillation}

Magic-state distillation is a standard fault-tolerant method for producing high-fidelity non-Clifford resource states from a larger number of noisy input magic-states~\cite{bravyi2005universal,reichardt2005quantum,bravyi2012low,meier2013magic,haah2017lowspace}. Its basic idea is to use only fault-tolerant Clifford operations, auxiliary measurements, and postselection to detect error patterns among imperfect resource states. A distillation protocol typically takes several noisy magic-states as input, encodes or processes them through a structured Clifford circuit, measures selected stabilizers or parity checks, and accepts the output only when the measurement outcomes indicate that no detectable error has occurred. Conditional on acceptance, the remaining output state has a lower error rate than the inputs. This many-to-one, postselected purification structure is illustrated in Fig.~\ref{fig:abc_schematic}(a).

The main advantage of magic-state distillation is that it can systematically suppress errors to very low levels and can therefore supply the high-fidelity $\Tstate$s required by large-scale fault-tolerant computation. Through concatenated protocols, block-code constructions, and optimized factory designs, distillation can reach target error rates far below those accessible by direct noisy preparation alone~\cite{jones2013multilevel,fowler2013surface,hastings2018sublog,ogorman2017realistic,haah2021measurement}. Its main drawback is resource overhead: improving output fidelity usually requires many input states, repeated rounds of checking and postselection, and substantial space-time cost in the surrounding fault-tolerant architecture. In surface-code settings, this overhead is typically realized as dedicated magic-state factories that consume encoded noisy inputs and produce purified logical $\Tstate$s for later injection into the computation~\cite{litinski2019magic,litinski2019notcostly}. For this reason, magic-state distillation is often regarded as the most mature and reliable route to high-fidelity magic-state preparation, but also the most resource-intensive one.

\subsection{Magic-state cultivation}

Magic-state cultivation is a more recent approach to logical magic-state preparation that seeks to reduce the overhead associated with factory-style distillation~\cite{gidney2024cultivation,bombin2024postselection,chen2026efficient}. Instead of beginning with many noisy resource states and purifying them through a large multi-copy protocol, cultivation starts from a small encoded magic-state and progressively improves its protection through staged code growth, intermediate verification, and, when necessary, postselection or restart. In this sense, cultivation replaces repeated large-scale purification with a stepwise increase in code protection around a single developing resource state. The staged injection, cultivation, growth, and grafting picture is shown in Fig.~\ref{fig:abc_schematic}(b).

The main appeal of cultivation is its potential to lower the qubit and space-time overhead required for logical $\Tstate$ preparation, especially in regimes where full distillation factories may be unnecessarily costly. At the same time, failed growth and verification stages increase the expected per-accepted-output cost $\Vexp$, so restart behavior and implementation details remain important in assessing its practical efficiency. Recent developments, including $\mathbb{RP}^2$-based variants, have further improved the method and extended its explicitly reported performance into the $10^{-6}$--$10^{-9}$ output-error regime~\cite{gidney2024cultivation,chen2026efficient}.

\subsection{Code-switching}

Code-switching is another approach to logical magic-state preparation that exploits the fact that some color-code and higher-dimensional topological-code constructions admit broader transversal or otherwise protected gate sets than the two-dimensional surface code~\cite{bombin2006topological,bombin2015gauge,bravyi2013classification,kubica2015unfolding,campbell2017roads}. The basic idea is to temporarily transfer quantum information into an encoding in which the desired non-Clifford operation or resource state can be realized more naturally, and then map the state back to the code used for subsequent computation. In subsystem-code language, a closely related mechanism is gauge-fixing, where the set of measured gauge operators is changed so that the same physical qubits are reinterpreted as different effective codes with different transversal gate properties~\cite{bombin2015gauge,anderson2014fault}. A representative qRM-to-Steane switching construction is sketched in Fig.~\ref{fig:abc_schematic}(c).

The main appeal of code-switching is that it can avoid some of the large multi-copy overhead associated with distillation by carrying out the crucial preparation step in a more favorable encoding. In this way, the cost of repeated purification is partly replaced by the cost of code conversion, gauge-fixing, and the fault-tolerant transfer process itself. Practical implementations often combine code conversion with verification or postselection in order to suppress residual logical error. Recent examples include protocols based on 3D-to-2D color-code transitions and flag-based postselection, which provide explicit logical $\Tstate$ preparation schemes within this framework~\cite{daguerre2025code,butt2024fault}. More broadly, earlier work on concatenated codes, stacked codes, pieceable constructions, and color-code decoding also supports this approach by showing how universality can be recovered through code conversion, code-switching, or related fault-tolerant gate constructions~\cite{jochymoconnor2014concatenated,jochymoconnor2016stacked,yoder2016pieceable,landahl2011fault,delfosse2014decoding,paetznick2013universal}.

\section{Method}\label{sec:method}

This section defines how heterogeneous results are compared without removing the assumptions under which they were obtained. The purpose of the method is not to convert all protocols into a single universal score, but to preserve the convention of each source while extracting the quantities needed for controlled comparison. This is necessary because distillation, code-switching, and cultivation are under different code families, noise models, postselection rules, and cost units.

The comparison is therefore organized around presented protocol configurations. A protocol configuration means one reported or transparently reconstructable set of assumptions and numerical outputs from an original paper. For each configuration, we record the assumptions needed to interpret the result and the quantities needed to compare it with other configurations. Numerical comparisons are made directly only when the corresponding units and error definitions are compatible; otherwise, the data are used to identify regimes and trade-offs rather than to produce a universal ordering.

\subsection{Comparison logic}

We use the comparison logic to separate output sufficiency, overhead, and interpretability. Output sufficiency is represented by the output error $\epsout$; overhead is represented by the single-attempt cost $\Vcost$, expected cost $\Vexp$, peak footprint $\Qpeak$, and latency $D$; and interpretability is represented by reporting completeness. Probabilistic restarts are folded into $\Vexp$ through the success probability $\Psucc$, so $\Psucc$ is treated as an auxiliary reconstruction quantity rather than as an independent comparison axis.

\subsection{Comparison quantities}

The quantities below implement this logic and form the minimum set needed to connect single-state preparation to fault-tolerant use. The output error $\epsout$ determines whether a prepared $\Tstate$ can satisfy a target fidelity. The single-attempt cost $\Vcost$ determines the raw overhead of one preparation attempt, while the expected cost $\Vexp$ records the restart overhead required to obtain one accepted output. The peak footprint $\Qpeak$ and latency $D$ describe how the preparation routine occupies hardware and time. Reporting completeness prevents missing assumptions from being interpreted as favorable performance.

\paragraph{Qubit footprint.}
The purpose of $\Qpeak$ is to describe the largest simultaneous qubit demand of a preparation routine. We record it in the original paper's footprint unit. When reporting physical qubits, $\Qpeak$ is treated as a physical-qubit footprint. When reporting only a logical-level construction, $\Qpeak$ is treated as the peak logical footprint or the corresponding protected-unit count used by that paper. Because physical and logical footprints cannot be converted without architecture-dependent assumptions, we do not force such a conversion unless the original paper provides one. Thus, direct numerical comparison of $\Qpeak$ is made only between compatible footprint units.

\paragraph{Latency.}
The purpose of $D$ is to record the duration of one preparation attempt in the time convention used by the original paper. In surface-code settings this may be presented in code cycles; in other constructions it may be presented in protection rounds, verification rounds, or another step. We therefore treat $D$ as a native duration rather than as a wall-clock time. Direct comparison of latency is made only when the underlying time baseline is compatible.

\paragraph{Single-attempt cost.}
The purpose of $\Vcost$ is to record the qubit-time cost of one preparation attempt before conditioning on acceptance. It is one of the two cost variables used in the family-level comparison. If a source reports only $\Vexp$ and does not provide the auxiliary information needed to reconstruct $\Vcost$, then $\Vcost$ is marked as unavailable rather than inferred.

\paragraph{Expected cost.}
The purpose of $\Vexp$ is to record the average qubit-time cost per accepted output. It is the main cost variable used in the cost--error regime map and in the algorithm-level serial-cost estimate. For probabilistic protocols, $\Vexp$ includes the restart overhead associated with failed preparation attempts.

\paragraph{Output error.}
The purpose of $\epsout$ is to locate the fidelity regime reached by each preparation route. We record the presented error rate, infidelity, or failure probability of the prepared logical $\ket{T}$ state. Because original papers do not always use identical error definitions, numerical comparison of $\epsout$ is made only when the definitions are compatible. Otherwise, $\epsout$ is retained as a native indicator of output quality.

\paragraph{Yield.}
Yield is used only as the auxiliary success probability $\Psucc$ connecting $\Vcost$ and $\Vexp$:
\begin{equation}
    \Vexp = \frac{\Vcost}{\Psucc},
    \qquad
    \Vcost = \Vexp\,\Psucc .
\end{equation}
Thus, when a source reports $\Vexp$ together with $\Psucc$, we reconstruct $\Vcost$; when it reports $\Vcost$ together with $\Psucc$, we reconstruct $\Vexp$. We do not use $\Psucc$ as an independent comparison dimension in the later figures, because $\Vcost$, $\Vexp$, and $\Psucc$ are constrained by the equation above.

\paragraph{Specification completeness.}
The purpose of this field is to record whether a presented configuration contains the assumptions needed for interpretation. These assumptions include the physical error model, code family, decoder or simulation convention, postselection rule, footprint, latency, the information needed to interpret $\Vcost$ and $\Vexp$, and output-error definition. This field is not a performance metric. It is included to prevent missing information from being treated as favorable performance.

\section{Numerical Results}\label{sec:results}

This section applies the comparison logic defined in \Cref{sec:method} to the presented protocol configurations. The purpose is to identify which trade-offs are supported by the current literature while keeping the original assumptions visible. We therefore present two complementary results.

The first result is a cost--error regime map. It compares the reported or reconstructed expected cost $\Vexp$ and output error $\epsout$ of the included configurations while preserving the cost convention of each paper. This result is used to locate the low-cost, intermediate-fidelity, and deep-fidelity regimes occupied by distillation, code-switching, and cultivation.

The second result is a family-level summary. It compares the same evidence across $\Vcost$, $\Vexp$, $\Qpeak$, $D$, $\epsout$, and reporting completeness. This result is used to show the strengths and limitations of each protocol family across several reported quantities, rather than to assign a single universal ranking.

\subsection{Cost--error regime map}

\begin{figure}[t]
\centering
\includegraphics[width=0.92\columnwidth]{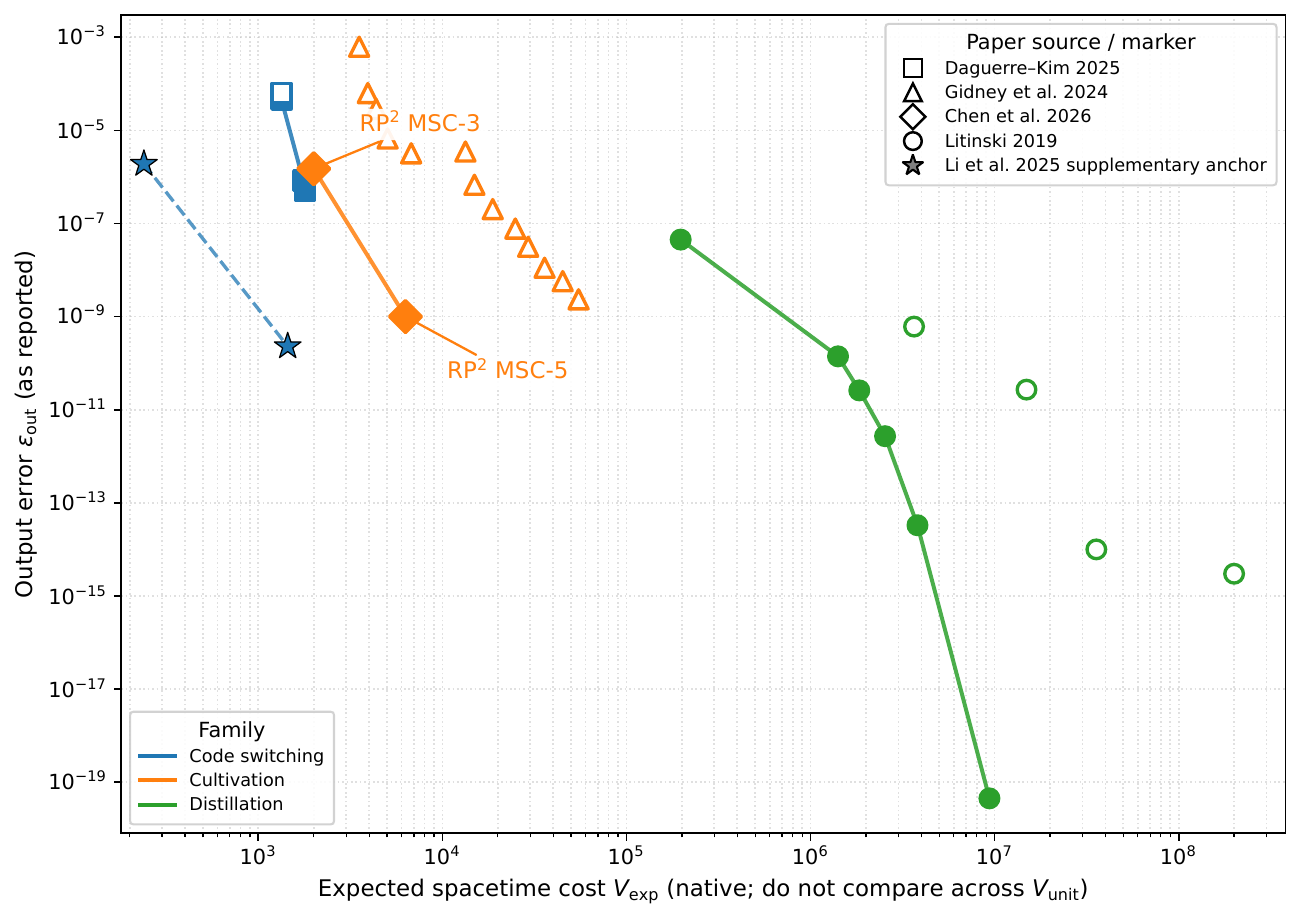}
\caption{Expected cost $\Vexp$ versus reported output error $\epsout$ for the included data rows. Marker shape indicates paper provenance, and filled markers denote within-family Pareto-frontier points in the displayed plane. Because the native cost unit differs across original papers, the horizontal axis should be read as a regime map rather than as a universal cross-paper ruler.}
\label{fig:pareto}
\end{figure}

The purpose of the regime map is to show how the expected per-accepted-output preparation cost changes with output fidelity across the presented configurations. \Cref{fig:pareto} therefore plots the expected cost $\Vexp$ against the output error $\epsout$. Each value is kept in the cost convention of its original paper, so the horizontal axis should be read as a native regime indicator rather than as a universal cross-paper cost ruler.

The plotted points are included only when the relevant assumptions and numerical quantities are explicitly represented or transparently reconstructable, as described in Appendix~\Cref{sec:app:reconstruction}. Litinski's distillation points are surface-code lattice-surgery factories. Daguerre and Kim's code-switching points use a qRM-to-Steane transition, equivalently a three-dimensional color-code to two-dimensional color-code-switching protocol. The Gidney--Shutty--Jones cultivation points start from a small color-code seed, grow through color-code cultivation, and escape into a matchable code used in a surface-code-style workflow. The Chen \emph{et al.} points are $\mathbb{RP}^2$ cultivation configurations ending in a rotated surface code.

Even under these restrictions, \Cref{fig:pareto} already reveals a clear hierarchy in the overhead regime. The code-switching rows occupy the lowest reported native single-attempt-cost region in the dataset, around $1.3 \times 10^{3}$--$1.8 \times 10^{3}$, with reported output errors in the approximate range $10^{-5}$--$10^{-7}$. This identifies code-switching as the strongest current option when the primary objective is low preparation overhead. However, its limitation is also clear: the tabulated rows do not yet reach the very deep output-error tail. In other words, code-switching is highly competitive as an engineering-efficient solution, but its presently existing protocol outcomes do not yet demonstrate the same universality for ultra-high-fidelity targets as the best distillation factories.

The cultivation rows now split into two visible bands. The original Gidney--Shutty--Jones magic-state-cultivation data correspond to an end-to-end protocol that starts by injecting \(\ket{T}\) into a distance-3 color code, cultivates it through growing color-code patches to a target fault distance (equivalently, code distance at the end of cultivation) of 3 or 5, and then escapes by grafting into a distance-15 matchable code. Across these end-to-end protocol outcomes, the displayed single-attempt costs are roughly \(3.5 \times 10^{3}\) to \(5.5 \times 10^{4}\), with displayed output errors ranging from about \(6 \times 10^{-4}\) down to \(2 \times 10^{-9}\) depending on the postselection cutoff~\cite{gidney2024cultivation}. The $\mathbb{RP}^2$ cultivation rows reported by Chen \emph{et al.} sharpen this picture by contributing two explicit end-to-end points at about $2.0 \times 10^{3}$ with $\epsout \approx 1.5 \times 10^{-6}$ and $6.3 \times 10^{3}$ with $\epsout \approx 1 \times 10^{-9}$, together with footprints of $247$ and $251$ qubits~\cite{chen2026efficient}. This pulls the cultivation frontier substantially leftward in the practically relevant $10^{-6}$--$10^{-9}$ regime. Relative to distillation, cultivation therefore looks materially stronger than before as a low-overhead route to useful logical $\ket{T}$ states.

The distillation rows span the highest native-cost region in the set and also the deepest reported output-error tail. In \Cref{fig:pareto}, the family-colored connector is a Pareto-frontier guide rather than a regression fit. Distillation is therefore the most conservative route when the design target is the smallest $\epsout$, but that advantage is obtained at substantially larger resource cost. Thus, distillation is not disfavored in general; it is disfavored only under an objective function dominated by low native cost, small footprint, and short preparation latency.

The regime map is used to separate two notions of performance. If the objective is low single-attempt cost within the presented configurations, code-switching occupies the strongest region, cultivation follows, and distillation is the most costly. If the objective is the lowest reported output error, the ordering reverses: distillation reaches the deepest error regime, while code-switching and cultivation remain limited by the currently tabulated evidence. This distinction is the main conclusion supported by \Cref{fig:pareto}.

This conclusion is restricted to the architectures represented in the included data. The comparison does not incorporate concatenated coding, multi-round hierarchical distillation beyond the represented points, or unreported larger-distance extrapolations. Including such constructions could change the relative standing of the protocol families.

The plotted provenance is as follows. Blue squares are the Daguerre--Kim code-switching rows from Tables~I and VII of Ref.~\cite{daguerre2025code}; orange triangles are the Gidney--Shutty--Jones cultivation data digitized from Fig.~1 of Ref.~\cite{gidney2024cultivation}, with discard information linked from Fig.~2 of the same reference where explicit; orange diamonds are the $\mathbb{RP}^2$ cultivation rows from Chen \emph{et al.}, using Table~I and Fig.~4(b) of Ref.~\cite{chen2026efficient}; and green circles are the Litinski distillation rows from Table~I of Ref.~\cite{litinski2019notcostly}. The starred dashed connector is a figure-only supplementary anchor from Table~II of Ref.~\cite{li2025transversal} and is excluded from the family-level summary.

\subsection{Family-level comparison}

\begin{figure}[t]
\centering
\includegraphics[width=0.90\columnwidth]{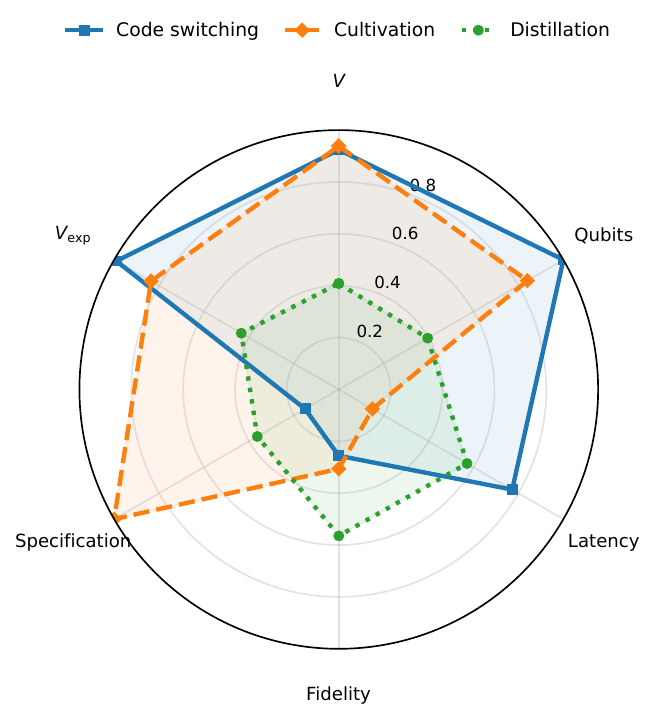}
\caption{Family-level summary over $\Vcost$, $\Qpeak$, $D$, $\epsout$, reporting completeness, and $\Vexp$. Each spoke reports the median normalized value over rows that provide the corresponding field; lower raw values are better for all quantitative resource/error fields.}
\label{fig:radar}
\end{figure}

The purpose of the family-level comparison is to show which advantages remain visible after the point-by-point data are grouped by preparation route. \Cref{fig:radar} summarizes the rows shown in \Cref{fig:pareto} across $\Vcost$, $\Qpeak$, $D$, $\epsout$, reporting completeness, and $\Vexp$. The plot is used as a visual summary of presented quantities, not as a single numerical score.

The summary includes only points in the core dataset. The starred supplementary anchors are excluded~\cite{litinski2019notcostly,daguerre2025code,gidney2024cultivation,chen2026efficient}. The Li \emph{et al.}, Vaknin \emph{et al.}, and Claes results are therefore treated as contextual extensions rather than family-level summary rows. In particular, Table~II of Ref.~\cite{li2025transversal} suggests a possible continuation of code-switching toward the $10^{-10}$ regime with higher native cost and additional protection, while distillation remains the only included family that reaches the much deeper $10^{-14}$--$10^{-20}$ regime. The nonzero floor in the radar plot marks missing or unavailable fields and should not be interpreted as a performance score.

For code-switching, the family-level summary is used to test whether the low-cost position seen in \Cref{fig:pareto} is accompanied by other low-overhead indicators. The blue polygon is widest on the direct-overhead axes $\Vcost$, $\Qpeak$, and $D$, and it also remains competitive on $\Vexp$. This supports the interpretation that code-switching is the strongest low-overhead family in the reported rows. Its $\epsout$ spoke is not dominant, however, which reflects the absence of tabulated code-switching rows in the deepest output-error regime.

For cultivation, the family-level summary is used to determine whether the newer $\mathbb{RP}^2$ data change the status of the family. The orange polygon is strengthened by the added rows because they provide explicit end-to-end footprints and low-cost points in the $10^{-6}$--$10^{-9}$ range. This moves cultivation closer to the low-overhead region than distillation. Its limitation is that latency and other implementation-level quantities are still presented less uniformly than for the other families, so the family remains harder to compare on all axes.

For distillation, the family-level summary is used to separate output-error strength from overhead cost. The green polygon is narrow on $\Vcost$, $\Vexp$, $\Qpeak$, and $D$ but extends furthest on $\epsout$. This confirms that distillation is the most mature route for very high-fidelity $\Tstate$ production, while also being the most expensive family in the presented low-overhead axes. Thus, distillation is not weak in principle; it is optimized for a different regime.

\subsection{Discussion}

The purpose of the discussion is to connect the two numerical summaries to the design choice they imply. The main distinction is between single-state efficiency and deepest attainable fidelity. A protocol family can be favorable under one criterion and unfavorable under the other.

Under a low-overhead criterion, the present data favor code-switching. This conclusion follows from the combination of low $\Vcost$, favorable $\Vexp$, small footprint, and short reported latency in the included data. The limitation is that the same rows do not extend to the deepest output-error regime. Code-switching is therefore best supported as a low-overhead preparation route, not yet as a complete replacement for distillation at ultra-low target errors.

Under an intermediate criterion, cultivation is the most changed family relative to earlier comparisons. The $\mathbb{RP}^2$ rows provide explicit end-to-end footprint information and move cultivation into a more competitive low-cost region. This supports the view that cultivation is no longer merely a speculative intermediate route. Its present limitation is reporting unevenness, especially in directly comparable latency and larger-distance data.

Under a stringent low-error criterion, distillation remains the strongest family. Its expected cost and footprint are larger, but it reaches the deepest output-error regime among the included rows. This explains why distillation can rank poorly in a low-overhead comparison while still being the most reliable choice when the target error is extremely small.

\section{Illustrative algorithm-level error-budget check}\label{sec:shor_case}

This section gives an illustrative error-budget check rather than a full resource estimate for Shor's algorithm. Its purpose is to show how a protocol that is attractive for preparing a single logical $\Tstate$ can fail to meet the much smaller per-state error target required by a large algorithmic workload. We use Shor factoring only as a familiar workload model with a large non-Clifford demand~\cite{shor1994algorithms,gidney2021factor}. The estimates below deliberately ignore scheduling, correlated logical faults, decoding latency, routing, factory placement, and wall-clock calibration across different time units.

\subsection{From modular arithmetic to logical T-state demand}

In optimized fault-tolerant implementations of Shor's algorithm, the dominant non-Clifford cost comes from modular exponentiation rather than from the quantum Fourier transform. Gidney and Eker\aa{} estimate that factoring an $n$-bit RSA integer requires approximately
\begin{align}
Q_{\mathrm{data}}(n) &\approx 3n + 0.002\,n\log_2 n, \\
N_{\mathrm{Tof}}(n) &\approx 0.3n^3 + 0.0005\,n^3\log_2 n, \\
L_{\mathrm{meas}}(n) &\approx 500n^2 + n^2\log_2 n,
\end{align}
where $Q_{\mathrm{data}}$ denotes the logical data-qubit count, $N_{\mathrm{Tof}}$ the Toffoli count, and $L_{\mathrm{meas}}$ the measurement depth in the abstract circuit model~\cite{gidney2021factor}. To connect these algorithm-level quantities to logical $\ket{T}$ preparation, the non-Clifford arithmetic must then be expressed in Clifford+$T$ form, where different Toffoli decompositions and $T$-depth optimizations lead to different effective $T$-state demands~\cite{selinger2013quantum,jones2013toffoli,gidney2018halving}.

To map algorithm-level non-Clifford workload to logical $\Tstate$ demand, the only gate-level input needed in the present resource estimate is the effective number of consumed $T$ states per Toffoli-equivalent operation. We therefore parameterize the compilation by a coefficient $c_T$ and evaluate two representative scenarios, $c_T=4$ and $c_T=7$. These values are not thresholds: $c_T=7$ represents the standard exact Clifford+$T$ Toffoli decomposition, while $c_T=4$ represents an optimized relative-phase or measurement-assisted Toffoli-equivalent accounting often used as a lower-overhead arithmetic scenario. The ratio between them is only $7/4$, so the purpose is a sensitivity check rather than a claim that the two cases define qualitatively different regimes. Under this convention, the total logical $\ket{T}$ demand is written as
\begin{equation}
N_T(n;c_T) \approx c_T\,N_{\mathrm{Tof}}(n),\qquad c_T\in\{4,7\},
\label{eq:Nt_from_toffoli}
\end{equation}
where each logical $T$ gate is assumed to consume one logical $\ket{T}$ state in an injection-based fault-tolerant implementation~\cite{jones2013toffoli,gidney2018halving}.

\begin{table}[t]
\caption{Shor-level non-Clifford demand used in the case study. Here $n$ is the RSA modulus size in bits, $c_T$ is the compilation-level coefficient converting Toffoli demand into logical $T$-state demand, $N_T$ is the resulting number of logical $T$ states consumed, and $\eta_T$ is the fraction of the total algorithm failure budget allocated to logical $\ket{T}$ injections. The RSA-2048 column uses $n=2048$ and the illustrative choice $\eta_T=10^{-2}$.}
\label{tab:shor_budget}
\small
\renewcommand{\arraystretch}{1.12}
\begin{tabular}{ll}
\toprule
\parbox[t]{0.37\columnwidth}{Quantity} &
\parbox[t]{0.48\columnwidth}{Expression / RSA-2048 value} \\
\midrule
\parbox[t]{0.37\columnwidth}{Logical data qubits} &
\parbox[t]{0.48\columnwidth}{$3n + 0.002\,n\log_2 n \Rightarrow 6.19\times 10^3$} \\
\parbox[t]{0.37\columnwidth}{Toffoli count $N_{\mathrm{Tof}}$} &
\parbox[t]{0.48\columnwidth}{$0.3n^3 + 0.0005\,n^3\log_2 n \Rightarrow 2.62\times 10^9$} \\
\parbox[t]{0.37\columnwidth}{$T$ count $N_T$ ($c_T=4$ scenario)} &
\parbox[t]{0.48\columnwidth}{$4N_{\mathrm{Tof}} \Rightarrow 1.05\times 10^{10}$} \\
\parbox[t]{0.37\columnwidth}{$T$ count $N_T$ ($c_T=7$ scenario)} &
\parbox[t]{0.48\columnwidth}{$7N_{\mathrm{Tof}} \Rightarrow 1.84\times 10^{10}$} \\
\parbox[t]{0.37\columnwidth}{Per-$T$ output-error target} &
\parbox[t]{0.48\columnwidth}{$\eta_T/N_T \Rightarrow 9.53\times 10^{-13}$ (4$T$), $5.44\times 10^{-13}$ (7$T$)} \\
\parbox[t]{0.37\columnwidth}{Measurement depth} &
\parbox[t]{0.48\columnwidth}{$500n^2 + n^2\log_2 n \Rightarrow 2.14\times 10^9$} \\
\bottomrule
\end{tabular}
\end{table}

\subsection{An algorithm-level error-budget filter}

Suppose that protocol $j$ produces logical $\ket{T}$ states with reported output error $\epsilon_{\mathrm{out},j}$. If the full algorithm consumes $N_T$ such states, then a first-order condition
for supplying all required $\ket{T}$ states directly from protocol $j$, without additional rounds of distillation or other post-processing, is
\begin{equation}
N_T(n;c_T)\,\epsilon_{\mathrm{out},j} \lesssim \eta_T,
\label{eq:error_budget}
\end{equation}
where $\eta_T$ is the portion of the total algorithm failure budget assigned to $\ket{T}$-injection faults. Throughout this section we take the illustrative choice $\eta_T=10^{-2}$, so that one percent of the total failure budget is reserved for all logical $T$-state consumptions in the algorithm.

This condition is a first-order union-bound filter. It should not be read as a complete fault-tolerant resource model, but as a conservative way to test whether a reported single-state output error is even in the right regime for a large workload.

For RSA-2048, Table~\ref{tab:shor_budget} implies a required per-state output error on the order of $10^{-12}$. This immediately changes the interpretation of the single-state comparison. In the present dataset, the best code-switching rows lie near $\epsout\sim 10^{-7}$, while the strongest explicit $\mathbb{RP}^2$ cultivation row reaches $\epsout=10^{-9}$~\cite{daguerre2025code,chen2026efficient}. Both remain several orders of magnitude above the RSA-2048 target. By contrast, the Litinski distillation ladder includes rows at $\epsout=3.3\times 10^{-14}$ and below~\cite{litinski2019notcostly}, which satisfy this illustrative budget. Indeed, the nearby distillation row at $2.7\times 10^{-12}$ is already slightly too noisy under the same accounting, so the effective threshold falls within the distillation family itself.

It is also useful to translate this budget condition into a measure of algorithmic reach. Define
\begin{equation}
n_{\max,j}(c_T,\eta_T)=\max\!\left\{n:\,N_T(n;c_T)\,\epsilon_{\mathrm{out},j}\le \eta_T\right\},
\label{eq:nmax}
\end{equation}
that is, the largest RSA modulus size supportable by protocol $j$ under the chosen $T$-state error budget. In this form, single-state fidelity becomes a direct ceiling on problem size. We will use this quantity below to compare representative protocol outcomes across the three protocol families.

\subsection{Resource implications for the three protocol families}

For a representative protocol outcomes $j$ with per-output expected cost $V_{\mathrm{exp},j}$, peak footprint $Q_{\mathrm{peak},j}$, and reported latency $D_j$, the serial algorithm-level supply cost may be approximated as
\begin{equation}
C^{\mathrm{serial}}_j(n;c_T) \approx N_T(n;c_T)\,V_{\mathrm{exp},j}.
\label{eq:serial_cost}
\end{equation}
If $m_j$ factories are run in parallel, a first-order peak-resource model is
\begin{equation}
Q^{\mathrm{tot}}_j(n;m_j) \approx Q_{\mathrm{data}}(n) + m_j\,Q_{\mathrm{peak},j}.
\label{eq:parallel_qubits}
\end{equation}
When the algorithm time baseline and the factory time baseline are directly comparable, one may also estimate the required factory multiplicity from $D_j$ and $L_{\mathrm{meas}}(n)$. In the present comparison, however, latency is reported in source-native units. Accordingly, Eq.~(\ref{eq:serial_cost}) should be read as a native-unit regime indicator, and Eq.~(\ref{eq:parallel_qubits}) as a footprint template rather than a universal wall-clock conversion.

With that caveat in place, Table~\ref{tab:shor_mapping} compares three representative protocol outcomes from \Cref{sec:results} under a common RSA-2048 resource mapping: the Daguerre--Kim postselected code-switching point, the Chen \emph{et al.} $\mathbb{RP}^2$ MSC-5 end-to-end cultivation point, and the Litinski distillation point at $\epsout=3.3\times10^{-14}$~\cite{daguerre2025code,chen2026efficient,litinski2019notcostly}. These points are not chosen by a single ranking criterion, but rather to illustrate three distinct reference cases. The code-switching point is included as a particularly low-cost protocol outcomes, the cultivation point as the strongest explicit end-to-end cultivation result for which footprint data are available, and the distillation point as the first core distillation protocol outcomes that clearly satisfies the RSA-2048 budget adopted here.

\begin{table*}[!htbp]
\caption{Representative Shor mappings for the three protocol families. Here $n_{\max}$ is defined by Eq.~(\ref{eq:nmax}) with $\eta_T=10^{-2}$. Peak footprint and latency are reported only in source-native units and are included as auxiliary engineering descriptors rather than as universal cross-paper scalars.}
\label{tab:shor_mapping}
\scriptsize
\renewcommand{\arraystretch}{1.12}
\begin{tabular}{llllll}
\toprule
\parbox[t]{0.15\textwidth}{Protocol family} &
\parbox[t]{0.08\textwidth}{\centering $\epsout$} &
\parbox[t]{0.10\textwidth}{\centering Peak footprint\\ (native)} &
\parbox[t]{0.09\textwidth}{\centering Latency\\ (native)} &
\parbox[t]{0.09\textwidth}{\centering $n_{\max}$\\ (4$T$/7$T$)} &
\parbox[t]{0.33\textwidth}{Implication for RSA-2048} \\
\midrule
\parbox[t]{0.15\textwidth}{Code-switching} &
\parbox[t]{0.08\textwidth}{\centering $5.1\times10^{-7}$} &
\parbox[t]{0.10\textwidth}{\centering 38} &
\parbox[t]{0.09\textwidth}{\centering 81} &
\parbox[t]{0.09\textwidth}{\centering 25\\21} &
\parbox[t]{0.33\textwidth}{Lowest single-$\Tstate$ overhead regime and smallest explicit footprint, but far too noisy to serve as a standalone RSA-2048 supplier.} \\

\parbox[t]{0.15\textwidth}{Cultivation} &
\parbox[t]{0.08\textwidth}{\centering $1.0\times10^{-9}$} &
\parbox[t]{0.10\textwidth}{\centering 251} &
\parbox[t]{0.09\textwidth}{\centering --} &
\parbox[t]{0.09\textwidth}{\centering 201\\167} &
\parbox[t]{0.33\textwidth}{Moves closer to the Shor regime while remaining compact, but still does not close the RSA-2048 budget on its own.} \\

\parbox[t]{0.15\textwidth}{Distillation} &
\parbox[t]{0.08\textwidth}{\centering $3.3\times10^{-14}$} &
\parbox[t]{0.10\textwidth}{\centering 39100} &
\parbox[t]{0.09\textwidth}{\centering 97.5} &
\parbox[t]{0.09\textwidth}{\centering 6277\\5209} &
\parbox[t]{0.33\textwidth}{First clearly RSA-2048-feasible point; much larger in footprint, but sufficient to close the adopted Shor error budget.} \\
\bottomrule
\end{tabular}
\end{table*}

For RSA-2048, the per-$T$ error target in Table~\ref{tab:shor_budget} is
\[
\epsilon_T^\star = 9.53\times 10^{-13}\ \text{(4$T$)},\quad
5.44\times 10^{-13}\ \text{(7$T$)}.
\]
For the three representative rows in Table~\ref{tab:shor_mapping}, the output errors are
\begin{align}
    \epsilon_{\mathrm{out}}^{\mathrm{CS}}=5.1\times10^{-7},\notag\\
    \epsilon_{\mathrm{out}}^{\mathrm{Cult}}=1.0\times10^{-9},\notag\\
    \epsilon_{\mathrm{out}}^{\mathrm{Dist}}=3.3\times10^{-14}.\notag
\end{align}
Equivalently, relative to the RSA-2048 per-$T$ target, these correspond to
\[
\frac{\epsilon_{\mathrm{out}}^{\mathrm{CS}}}{\epsilon_T^\star}
\approx 5.35\times10^{5}\ \text{(4$T$)},\ 9.38\times10^{5}\ \text{(7$T$)},
\]
\[
\frac{\epsilon_{\mathrm{out}}^{\mathrm{Cult}}}{\epsilon_T^\star}
\approx 1.05\times10^{3}\ \text{(4$T$)},\ 1.84\times10^{3}\ \text{(7$T$)},
\]
\[
\frac{\epsilon_{\mathrm{out}}^{\mathrm{Dist}}}{\epsilon_T^\star}
\approx 3.46\times10^{-2}\ \text{(4$T$)},\ 6.07\times10^{-2}\ \text{(7$T$)}.
\]
Under the illustrative RSA-2048 budget adopted here, the selected distillation row lies within the standalone-supplier regime, while the representative code-switching and cultivation rows do not. The same conclusion is reflected by the reach metric $n_{\max}$: the representative rows support only $n_{\max}=25$ (4$T$) or $21$ (7$T$) for code-switching, $201$ (4$T$) or $167$ (7$T$) for cultivation, and $6277$ (4$T$) or $5209$ (7$T$) for distillation.

A more informative comparison is obtained by anchoring the resource estimate at the largest RSA size that the representative code-switching point can support at $p=10^{-3}$. Under the $c_T=4$ compilation scenario, Eq.~(\ref{eq:nmax}) gives $n_{\max}^{\mathrm{CS}}=25$, for which $N_T(25;4)\approx 1.89\times10^{4}$ and the required per-$T$ target is $\epsilon_T^\star(25)\approx 5.29\times10^{-7}$. Using the same representative rows as above, the resulting serial total preparation costs are approximately $2.98\times10^{7}$ for code-switching, $1.19\times10^{8}$ for cultivation, and $7.20\times10^{10}$ for distillation in their respective native units. Thus, at the largest RSA size supported by the representative code-switching point itself, code-switching is also the least expensive standalone supplier in total serial cost, cultivation is the next closest low-overhead alternative, and distillation remains the highest-overhead but highest-fidelity option.

This case study therefore sharpens a distinction that was already present in \Cref{sec:results}. Code-switching remains the best-supported \emph{single-state} engineering option when low single-attempt cost, small footprint, and short explicit latency are the main priorities. Cultivation remains a meaningfully strengthened intermediate family whose best explicit row moves closer to, but still does not reach, cryptographically relevant Shor fidelity. Distillation is still the least attractive family from the standpoint of single-state overhead, yet it becomes the most robust choice once a full algorithm-level error budget is imposed. In other words, the protocol family that offers the best overall balance for preparing a single logical $\ket{T}$ state need not be the best standalone backend for large-scale Shor factoring. That mismatch is precisely what makes the hybrid strategies already suggested in \Cref{sec:results} look especially natural at the level of complete algorithms.

\section{Scope and limitations}\label{sec:limitations}

The purpose of this section is to state the conditions under which the comparison should be interpreted. The results are limited by the presented data available in the current literature. Only original papers with explicit or transparently reconstructable numerical data are included in the quantitative comparison; other logical-$\Tstate$ preparation routes remain contextual.

The main limitation is that the included papers do not use a common presenting convention. Cost, latency, footprint, and output error are not always defined in the same units or under the same physical assumptions. For this reason, \Cref{fig:pareto} is presented in native units, and \Cref{fig:radar} is used only as a normalized family-level visualization. These figures support regime-level conclusions, but they do not establish a universal numerical ranking across all possible implementations.

A second limitation is uneven reporting completeness across protocol families. The $\mathbb{RP}^2$ cultivation paper improves the evidence for cultivation by providing explicit end-to-end footprint and expected-cost information. However, directly comparable latency remains unavailable across parts of the cultivation family, and some quantities elsewhere still require digitization or partial reconstruction. Thus, cultivation can now be judged on firmer low-overhead evidence, but it still cannot be compared with code-switching and distillation on every implementation-level quantity with the same completeness.

These limitations bound the claims rather than invalidating the comparison. The paper compares the presented configurations that satisfy the stated inclusion rule. It does not claim a final ranking of all logical $\Tstate$ preparation methods.

Future comparisons should therefore target the missing information directly. Cultivation would benefit most from directly comparable latency tables and additional larger-distance configurations, since the $\mathbb{RP}^2$ results already improve the family's footprint evidence. Code-switching and cultivation both need directly tabulated configurations in the $10^{-10}$ regime and below, so that their deeper-fidelity behavior can be compared with distillation on equal footing. More consistent definitions of output error, cost, and latency across papers would also allow sharper Pareto statements than the present native-unit regime map can support.

\section{Conclusion}\label{sec:conclusion}

This paper compared three representative families of logical $\Tstate$ preparation protocols, namely magic-state distillation, code-switching, and magic-state cultivation, using currently available results. Rather than reducing these heterogeneous protocols to a single aggregate score, we evaluated them across several explicit resource dimensions, including single-attempt cost, expected cost, qubit footprint, latency, output fidelity, and the extent of available reporting.

Within the results considered here, distillation remains the family that reaches the lowest output-error regime. Cultivation is substantially strengthened by the $\mathbb{RP}^2$ results, which provide explicit end-to-end footprints and lower-cost points in the $10^{-6}$ and $10^{-9}$ regimes, although latency reporting remains incomplete. Code-switching occupies the lowest cost range in the present dataset, while also showing the smallest reported footprint, short reported latency, and favorable $\Vexp$. These results suggest that the preferred family depends on the target objective: code-switching is favored under a low-overhead criterion within the reported rows, whereas distillation is favored under a stringent low-error criterion.

At the same time, this comparison remains limited by the current literature, which reports results under different code families, noise assumptions, decoder models, and cost conventions. It should therefore be read not as a final universal ranking, but as a structured comparison of the evidence currently available. Future work should target the missing data directly: code-switching needs explicitly tabulated larger-distance data to test whether its low-overhead advantage persists at lower output error; cultivation needs more uniformly reported latency and additional larger-distance configurations; and hybrid routes should be examined to determine whether code-switching or cultivation can supply a low-overhead initial state before distillation supplies deeper error suppression.

\section*{Acknowledgments}
We thank Ying Li and Guo Zhang for useful discussions. This work is supported by Beijing Natural Science Foundation Z250004, Quantum Science and Technology-National Science and Technology Major Project (2023ZD0300200), 
NSAF (Grant No.~U2330201), the National Natural Science Foundation of China Grant (No.~12361161602), 
Beijing Science and Technology Planning Project (Grant No.~Z25110100810000), and the High-performance Computing Platform of Peking University.

\appendix

\section{Measurement and data extraction}
\label{sec:app:reconstruction}

The purpose of the data extraction procedure is to ensure that every numerical row used in the comparison has a traceable paper and an interpretable set of assumptions. For each original paper, we include a presented configuration only when the relevant quantities are explicitly presented or can be reconstructed without adding assumptions not present in the paper.

Quantities presented directly in tables are transcribed as written. Quantities shown only in figures are included only when they can be read off with sufficient clarity and when the associated assumptions are explicit. Mixed table-plus-figure reconstruction is used only when the connection between the plotted point and the tabulated metadata is explicit. This is how the two Chen \emph{et al.} $\mathbb{RP}^2$ cultivation rows are added: Table~I of Ref.~\cite{chen2026efficient} fixes the end-to-end logical error rate, discard rate, final distance, and footprint, while Fig.~4(b) of the same reference supplies the corresponding native expected space-time cost. If a quantity is not reported, it is marked as missing rather than estimated from informal descriptions.

The provenance of the plotted rows is therefore paper-specific. In \Cref{fig:pareto}, the distillation circles come from Table~I of Ref.~\cite{litinski2019notcostly}; the code-switching squares come from Tables~I and VII of Ref.~\cite{daguerre2025code}; the cultivation triangles come from digitizing the Gidney--Shutty--Jones expected-cost curve in Fig.~1 of Ref.~\cite{gidney2024cultivation}, with discard data attached from Fig.~2 of the same reference where that linkage is explicit; and the cultivation diamonds come from Table~I plus Fig.~4(b) of Ref.~\cite{chen2026efficient}. The plotted dataset retains only starred supplementary anchors from Table~II of Ref.~\cite{li2025transversal}.

The extraction procedure also determines how missing latency is treated. All cost-like quantities are kept in the native unit of the original paper unless a transparent reconstruction of $\Vcost$ or $\Vexp$ from reported footprint, depth, and postselection/restart conventions is possible. For the Chen rows, latency is deliberately left unscored in the subset. Appendix~A of Ref.~\cite{chen2026efficient} gives a round-length convention for space-time accounting, but not a directly tabulated protocol-time baseline aligned with the other families. We therefore use the paper only for the fields that are explicit or cleanly digitizable. This is why the comparison distinguishes between a regime map and a normalized family-level summary instead of merging all configurations into a single ranking.

\section{Normalization for family-level visualization}
\label{sec:app:normalization}

The purpose of the normalization is to display several heterogeneous quantities on one family-level visual scale. The normalized values are used only for \Cref{fig:radar}. They are not used to define a universal performance score.

For each quantitative field, values are normalized within the dataset after applying a logarithm to scale-like quantities such as expected cost, footprint, latency, and output error. For a field where smaller values are better, the displayed score is
\begin{equation}
s_i = s_{\min} + (1-s_{\min})
\left[
1-\frac{\log x_i-\min_j \log x_j}{\max_j \log x_j-\min_j \log x_j}
\right].
\end{equation}
For specification-completeness fields, the corresponding non-logarithmic version is used. For fields where larger values are better, the bracketed term is used without the leading $1-$. The floor $s_{\min}$ is introduced only to keep missing or unavailable family-level fields visible in the plot. It is not interpreted as a performance score.

\bibliographystyle{quantum}
\bibliography{references}

\end{document}